\documentclass[12pt]{iopart}
\usepackage{blackboard}
\def\op#1{#1}
\def\ket#1{| #1 \rangle}
\def\bra#1{\langle #1 |}
\def\ip#1#2{\langle #1 \mid #2 \rangle}

\def\diag{\mbox{\rm diag}}
\def\A{{\cal A}}
\def\C{{\cal C}}
\def\H{{\cal H}}
\def\L{{\cal L}}

\def\dim{\mathop{\rm dim}}
\newtheorem{definition}{Definition}
\newtheorem{theorem}{Theorem} 
\newtheorem{lemma}{Lemma} 
\newcounter{exNo}
\newenvironment{example}{\refstepcounter{exNo}\par\parskip 2ex\noindent{\bf Example \arabic{exNo}:} \nobreak}{\rule{.6em}{0.6em}\par\parskip 2ex\noindent}
\newenvironment{proof}{\noindent{\sc Proof:} }{\rule{.6em}{0.6em}\par\parskip 2ex\noindent}
\begin{document}
\bibliographystyle{prsty}
\title{Criteria for reachability of quantum states}
\author{S G Schirmer and A I Solomon}
\address{Quantum Processes Group and Dept of Applied Maths, The Open University, \\
         Milton Keynes, MK7 6AA, United Kingdom}
\ead{\mailto{S.G.Schirmer@open.ac.uk},\mailto{A.I.Solomon@open.ac.uk}}
\author{J V  Leahy}
\address{Department of Mathematics and Institute of Theoretical Science, 
         University of Oregon, Eugene, Oregon, 97403, USA}
\ead{\mailto{leahy@math.uoregon.edu}}
\date{\today}
\begin{abstract}
We address the question of which quantum states can be inter-converted under the action
of a time-dependent Hamiltonian.  In particular, we consider the problem as applied to
mixed states, and investigate the difference between pure and mixed-state controllability
introduced in previous work.  We provide a complete characterization of the eigenvalue
spectrum for which the state is controllable under the action of the symplectic group. 
We also address the problem of which states can be prepared if the dynamical Lie group
is not sufficiently large to allow the system to be controllable.
\end{abstract}
\maketitle
\section{Introduction}

The subject of control of quantum systems has been a fruitful area of investigation
lately.  The growing interest in the subject can be attributed both to theoretical and
experimental breakthroughs that have made control of quantum phenomena an increasingly
realistic objective, as well as the prospect of many exciting new applications such as 
quantum computers \cite{IJMPA16p3335} or quantum chemistry \cite{SCI288p0824}, which 
attracts researchers from various fields.

Among the theoretical problems that have received considerable attention lately is the
issue of controllability of quantum systems.  Various aspects such as the controllability 
of quantum systems with continuous spectra \cite{IEEE39CDC2803, JMP24p2608}, wavefunction
controllability for bilinear quantum systems \cite{CRAS330p327, CP267p1, IEEE39CDC3003},
controllability of distributed systems \cite{PNAS94p4828}, controllability of molecular
systems \cite{PRA51p960}, controllability of spin systems \cite{qph0106115}, controllability
of quantum evolution in NMR spectroscopy \cite{MP96p1739}, and controllability of quantum
systems on compact Lie groups \cite{IEEE39CDC1086, JPA34p1679, PRA63n063410, qph0110147} 
have been addressed, and related problems such as the dynamical realizability of kinematical 
bounds on the optimization of observables \cite{IEEE39CDC3002, PRA63n025403}, and the 
relation between controllability and universality of quantum gates \cite{PRA54p1715}, as
well as the information-theoretic limits of control \cite{PRL84p1156} have been studied.

In this process, various notions of controllability have been introduced.  Recent work on
controllability of quantum systems on compact Lie groups has finally shown that the degree
of controllability of a quantum system depends on its dynamical Lie group, and that many 
different notions of controllability are in fact equivalent.  In particular, it has been 
proved that quantum systems evolving on a compact Lie group, such as closed quantum systems
with a discrete energy spectrum, are either density matrix / operator controllable, 
pure-state / wavefunction controllable, or not controllable \cite{qph0106128, qph0108114}.  
For density matrix, operator or completely controllable quantum systems, every kinematically 
admissible target state or operator can be dynamically realized, and the kinematical bounds
on the expectation values (ensemble averages) of observables are always dynamically attainable
\cite{PRA63n025403}.  Fortunately, many quantum systems have been shown to be completely
controllable \cite{JPA34p1679, PRA63n063410, qph0108114}. 

Nevertheless, there are quantum systems that are either only pure-state controllable or 
not controllable at all.  For instance, it has been shown that the dynamical Lie group 
of certain atomic systems with degenerate energy levels is the (unitary) symplectic group,
which corresponds to pure-state controllability \cite{qph0108114}.  Other systems with 
certain symmetries may be either pure-state controllable or non-controllable depending
on the symmetry.  For instance, given a system with $N$ equally spaced energy levels and
uniform dipole moments for transitions between adjacent levels, the dynamical Lie group
is the symplectic group if the dimension of its Hilbert space $N$ is even, but it is the
orthogonal group if $N$ is odd \cite{JPA35p2327}.  For these systems, the question of 
dynamical reachability of target states, which is important in many applications, remains.
In this paper, we address this problem by studying the action of the dynamical Lie group
of pure-state-only and non-controllable quantum systems on the kinematical equivalence 
classes of states.  Explicit criteria for dynamical reachability of states are derived 
for systems whose dynamical Lie group is the (unitary) symplectic group or the orthogonal
group.

\section{Quantum states and kinematical/dynamical equivalence classes}

We consider a quantum system whose state is represented by a density matrix acting on
a Hilbert space $\H$ of dimension $N$.  A density matrix always has a discrete spectrum
with non-negative eigenvalues $w_n$ that sum to one, $\sum_n w_n =1$, and a spectral 
resolution of the form
\begin{equation} \label{eq:DM}
  \op{\rho} = \sum_{n=1}^N w_n \ket{\Psi_n}\bra{\Psi_n},
\end{equation}
where $\ket{\Psi_n}$ are the eigenstates of $\op{\rho}$.  The $\ket{\Psi_n}$ for $1\le
n\le N$ are elements of the Hilbert space $\H$ and can always be chosen so as to form
a complete orthonormal set for $\H$.  The $\bra{\Psi_n}$ are the corresponding dual 
states defined by
\begin{equation}
  \ip{\Psi_n}{\Psi_m} = \delta_{mn} \qquad \forall m,n.
\end{equation}

Conservation laws such as conservation of energy and probability require the time 
evolution of any (closed) quantum system to be unitary.  Thus, given a Hilbert space
vector $\ket{\Psi_0}$, its time evolution is determined by $\ket{\Psi(t)}=\op{U}(t)
\ket{\Psi_0}$ where $\op{U}(t)$ a unitary operator for all $t$ and $\op{U}(0)=\op{I}$.
Hence, a density matrix $\op{\rho}_0$ must evolve according to 
\begin{equation}
  \op{\rho}(t)=\op{U}(t)\op{\rho}_0\op{U}(t)^\dagger,
\end{equation}
where $\op{U}(t)$ is unitary for all times.  This constraint of unitary evolution 
induces kinematical restrictions on the set of target states that are physically 
admissible from any given initial state.
\begin{definition}
Two quantum states represented by density matrices $\op{\rho}_0$ and $\op{\rho}_1$ are
\emph{kinematically equivalent} if there exists a unitary operator $\op{U}$ such that 
$\op{\rho}_1 = \op{U} \op{\rho}_0 \op{U}^\dagger$.
\end{definition}
Thus, the constraint of unitary evolution partitions the set of density matrices on
$\H$ into (infinitely many) kinematical equivalence classes.  It is well known that 
two density matrices $\op{\rho}_0$ and $\op{\rho}_1$ are kinematically equivalent if
and only if they have the same eigenvalues.  The kinematical equivalence classes are 
therefore determined by the eigenvalues of $\op{\rho}$.  
Furthermore, we introduce the following classification of density matrices according 
to their eigenvalues, which we shall relate to the degree of controllability of the 
system.
\begin{definition}[Classification of density matrices]
Every density matrix is of one of the following types. 
\begin{enumerate}
\item\label{type1} \emph{Completely random ensembles:}
      Density matrices whose spectrum consists of a single eigenvalue $w_1=\frac{1}{N}$
      that occurs with multiplicity $N$.
\item\label{type2} \emph{Pure-state-like ensembles:}
      Density matrices whose spectrum consists of two distinct eigenvalues, one of 
      which occurs with multiplicity one and the other with multiplicity $N-1$.
\item \label{type3} \emph{General ensembles:}
      Density matrices whose spectrum consists of at least two distinct eigenvalues, 
      at least one of which occurs with multiplicity $N_1$ where $2\le N_1\le N-2$; 
      or density matrices whose spectrum consists of $N$ distinct eigenvalues ($N\ge 2$).
\end{enumerate}
\end{definition}

Note that type (\ref{type2}) (pure-state-like ensembles) includes density matrices 
representing pure states such as $\op{\rho}=\mbox{diag}(1,0,0,0)$ but not every density
matrix in this class represents a pure state.  For instance, $\op{\rho}=\mbox{diag}
(0.7,0.1,0.1,0.1)$ is of type (\ref{type2}) but does \emph{not} represent a pure state.  

Given a specific quantum system with a control-dependent Hamiltonian of the form
\begin{equation} \label{eq:H}
  \op{H}[f_1(t),\ldots,f_M(t)] = \op{H}_0 + \sum_{m=1}^M f_m(t) \op{H}_m,
\end{equation}
where the $f_m$, $1\le m\le M$, are (independent) bounded measurable control functions,
the question arises which states are dynamically reachable from a given initial state.
Clearly, the set of potentially dynamically reachable states is restricted to states
within the same kinematical equivalence class as the initial state.  However, not every
kinematically admissible target state is necessarily dynamically reachable.  Since the 
time-evolution operator $\op{U}(t)$ has to satisfy the Schrodinger equation
\begin{equation} \label{eq:SE}
  \rmi\hbar \frac{d}{d t} \op{U}(t) = \op{H}[f_1(t),\ldots,f_M(t)] \op{U}(t),
\end{equation}
where $\op{H}$ is the Hamiltonian defined above, only unitary operators of the form
\begin{equation}
  \op{U}(t)=\exp_+\left\{-\frac{\rmi}{\hbar}\op{H}
                          \left[f_1(t),\ldots,f_M(t)\right]\right\},
\end{equation}
where $\exp_+$ denotes the time-ordered exponential, qualify as evolution operators.  
Using, for instance, the Magnus expansion of the time-ordered exponential, it can be 
seen that only unitary operators of the form $\exp(\op{x})$, where $\op{x}$ is an 
element in the dynamical Lie algebra $\L$ generated by the skew-Hermitian operators 
$\rmi\op{H}_0,\ldots,\rmi\op{H}_M$, are dynamically realizable.  These operators form
the dynamical Lie group $S$ of the system.  
\begin{definition}
Two kinematically equivalent states $\op{\rho}_0$ and $\op{\rho}_1$ are \emph{dynamically
equivalent} if there exists a unitary operator $\op{U}$ in the dynamical Lie group $S$ 
such that $\op{\rho}_1=\op{U}\op{\rho}_0\op{U}^\dagger$.
\end{definition}
This dynamical equivalence relation subdivides the kinematical equivalence classes.

In the following, we shall be particularly concerned with the unitary group $U(N)$, the
special unitary group $SU(N)$, the (unitary) symplectic group $Sp(\frac{N}{2})$ and the
(unitary) orthogonal group $SO(N)$.  As usual, the unitary group $U(N)$ is the compact 
Lie group consisting of all regular $N\times N$ matrices $\op{U}$ that satisfy 
$\op{U}^\dagger\op{U}=\op{U}\op{U}^\dagger =\op{I}$.  The special unitary group $SU(N)$ 
is the subgroup of $U(N)$ consisting of all unitary matrices $\op{U}\in U(N)$ whose 
determinant is $+1$.  For our purposes in this paper, we define the symplectic group
and the special orthogonal group as follows.

\begin{definition}
The (unitary) \emph{symplectic group} $Sp(\ell)$ is the subgroup of $SU(2\ell)$ 
consisting of all unitary operators of dimension $2\ell$ that satisfy 
$\op{U}^T\op{J}\op{U}=\op{J}$ for 
\begin{equation} \label{eq:Jsp}
  \op{J}= \left( \begin{array}{cc} 0 & \op{I}_\ell \\
                                 -\op{I}_\ell & 0 
                 \end{array} \right), 
\end{equation}
where $\op{I}_\ell$ is the identity matrix of dimension $\ell$.
\end{definition}

\begin{definition}
The (unitary) \emph{special orthogonal group} $SO(N)$ is the subgroup of $SU(N)$ 
consisting of all unitary operators of dimension $N$ that satisfy 
$\op{U}^T\op{J}\op{U}=\op{J}$ for 
\begin{equation} \label{eq:Jso}
  \op{J}= \left( \begin{array}{cc} 0 & \op{I}_\ell \\
                                 \op{I}_\ell & 0 
                 \end{array} \right), \quad N=2\ell, \quad 
  \op{J}= \left( \begin{array}{ccc} 1 & 0 & 0 \\
                                    0 & 0 & \op{I}_\ell \\
                                    0 & \op{I}_\ell & 0 
                 \end{array} \right), \quad N=2\ell+1. 
\end{equation}
\end{definition}

\section{Dynamical Lie group action on the kinematical equivalence classes}

The set of quantum states that is dynamically accessible from a given initial state 
$\op{\rho}_0$ depends on the action of the dynamical Lie group $S$ on the kinematical
equivalence classes of density operators.

\begin{definition}
The dynamical Lie group $S$ of a quantum system is said to \emph{act transitively} on 
a kinematical equivalence class $\C$ of density matrices if any two states in $\C$ are 
\emph{dynamically equivalent}.
\end{definition}
Since the equivalence class of completely random ensembles [type (\ref{type1}) above] 
consists only of a single state $\op{\rho}=\frac{1}{N}\op{I}_N$, it follows immediately
that \emph{every} group acts transitively on this equivalence class.   

Any dynamical Lie group $S$ that does not act transitively on the kinematical equivalence 
class of pure states, acts transitively only on the trivial kinematical equivalence class
of completely random ensembles.  Furthermore, from classical results by Montgomery and 
Samelson \cite{43Montgomery}, it follows that $U(N)$, $SU(N)$, $Sp(\frac{1}{2}N)$ and 
$Sp(\frac{1}{2}N)\times U(1)$ are the \emph{only} dynamical Lie groups (up to isomorphism) 
that act transitively on the equivalence class of pure states.  Therefore, any dynamical 
Lie group $S$ that is not isomorphic to either $U(N)$, $SU(N)$, $Sp(\frac{1}{2}N)$ or 
$Sp(\frac{1}{2}N)\times U(1)$ acts transitively only on type (\ref{type1}) states, i.e.,
completely random ensembles.  $U(N)$ and $SU(N)$ clearly act transitively on \emph{every}
kinematical equivalence class of states, which leaves only $Sp(\frac{1}{2}N)$ and 
$Sp(\frac{1}{2}N)\times U(1)$, whose action on the kinematical equivalence classes of 
states we shall now address.  

We begin by showing that transitive action of $Sp(\frac{1}{2}N)$ on pure states implies 
transitive action on all equivalence classes of type (\ref{type2}).  We shall prove this
result for the standard representation of $Sp(\frac{1}{2}N)$ as defined above.  To see 
that this is sufficient, note that lemma 4.2 in \cite{qph0106128} shows that whenever 
the dynamical Lie algebra of a quantum system of the type considered in this paper is 
\emph{isomorphic} to $sp(\frac{1}{2}N)$, then it is \emph{conjugate} to $sp(\frac{1}{2}N)$
via an element in $U(N)$.  Thus, if the dynamical Lie group $S$ of the system is of type 
$Sp(\frac{1}{2}N)$ then it is not only isomorphic to the standard representation of 
$Sp(\frac{1}{2}N)$, but there exists a unitary transformation (basis change) $\op{B}$ 
that maps any unitary operator in $\op{U}\in S$ to a unitary operator $\tilde{U}=\op{B}
\op{U}\op{B}^\dagger$ in the standard representation of $Sp(\frac{1}{2}N)$, i.e., $S$ is
unitarily equivalent to the standard representation of $Sp(\frac{1}{2}N)$.

Note that theorem 6 in \cite{qph0106128} gives a general condition for transitive action
of a dynamical Lie group $S\subset U(N)$ on a kinematical equivalence class of states 
represented by a density matrix $\op{\rho}$:  the action is transitive if and only if
\begin{equation} \label{eq:dimFormula}
  \dim U(N) - \dim S = \dim \C_\op{\rho} - \dim (\C_\op{\rho} \cap S),
\end{equation}
where $\C_\op{\rho}$ is the centralizer of $\op{\rho}$ and $\C_\op{\rho}\cap S$ is the
intersection of the centralizer with $S$.  However, since determination of the dimension 
of $\C_\op{\rho}$, and especially $\C_\op{\rho}\cap S$, tends to be very difficult in
practice (see \ref{appendix:B} for an example) we shall \emph{not} use this result but 
pursue an alternative approach instead.

\begin{lemma} \label{lemma1}
$Sp(\frac{1}{2}N)$ acts transitively on all kinematical equivalence classes of density
matrices whose eigenvalues satisfy $w_1 \neq w_2=w_3=\ldots=w_N$. 
\end{lemma}

\begin{proof}
Any $\op{\rho}$ with eigenvalues $w_1\neq w_2=w_3=\ldots=w_N$ can be written as
\[
  \op{\rho} = w_1 \ket{\Psi}\bra{\Psi} + w_2 \op{P}(\ket{\Psi}^\perp),
\] 
where $\op{P}(\ket{\Psi}^\perp)$ is the projector onto the orthogonal complement of the
subspace spanned by $\ket{\Psi}$.  Hence, any pair of kinematically equivalent states of
this type is of the form
\begin{eqnarray*}
 \op{\rho}_0 &=& w_1\ket{\Psi^{(0)}}\bra{\Psi^{(0)}}+w_2\op{P}(\ket{\Psi^{(0)}}^\perp)\\
 \op{\rho}_1 &=& w_1\ket{\Psi^{(1)}}\bra{\Psi^{(1)}}+w_2\op{P}(\ket{\Psi^{(1)}}^\perp).
\end{eqnarray*}
Since $Sp(\frac{1}{2}N)$ acts transitively on the equivalence class of pure states, there
exists a unitary operator $\op{U}\in Sp(\frac{1}{2}N)$ such that $\op{U}\ket{\Psi^{(0)}}=
\ket{\Psi^{(1)}}$.  Furthermore, $\op{U}$ automatically maps the orthogonal complement 
of $\ket{\Psi^{(0)}}$ onto the orthogonal complement of $\ket{\Psi^{(1)}}$ since it is
unitary and thus we have
\[
  \op{U}\op{\rho}^{(0)}\op{U}^\dagger 
 = w_1 \ket{\Psi^{(1)}}\bra{\Psi^{(1)}} + w_2 \op{P}(\ket{\Psi^{(1)}}^\perp)
 = \op{\rho}^{(1)}.
\]
Hence, $Sp(\frac{1}{2}N)$ acts transitively on all equivalence classes of density 
matrices whose eigenvalues satisfy $w_1\neq w_2=w_3=\ldots=w_N$.
\end{proof}
However, the action of $Sp(\frac{1}{2}N)$ on the class of pure states is not two-point
transitive as the following example shows.

\begin{example}\label{example:one}
Let $N=2\ell$ and $\vec{a}$ and $\vec{b}$ be two unit vectors in $\CC^N$.  Since
$N=2\ell$, we can partition the vectors as follows
\[
  \vec{a}=\left(\begin{array}{c}\vec{a}_1 \\ \vec{a}_2 \end{array}\right), \quad
  \vec{b}=\left(\begin{array}{c}\vec{b}_1 \\ \vec{b}_2 \end{array}\right), 
\]
where $\vec{a}_j$, $\vec{b}_j$ for $j=1,2$ are vectors in $\CC^\ell$.  Since $Sp(\ell)$
acts transitively on the unit sphere in $\CC^N$ it follows that there exists a $\op{U}
\in Sp(\ell)$ such that $\op{U}\vec{a}=\vec{b}$.  However, since any unitary operator 
in $Sp(\ell)$ satisfies $\op{U}^T\op{J}\op{U}=\op{J}$ with $\op{J}$ as in (\ref{eq:Jsp}),
we have $\op{U}=\op{J}^\dagger\op{U}^*\op{J}$ and thus $\op{J}^\dagger\op{U}^*\op{J}
\vec{a}=\vec{b}$ or equivalently $\op{U}\op{J}\vec{a}^* = \op{J}\vec{b}^*$.  Noting
that
\[
  \op{J}\vec{a}^*=\left(\begin{array}{c} -\vec{a}_2^* \\ 
                                          \vec{a}_1^* 
                        \end{array}\right), \quad
  \op{J}\vec{b}^*=\left(\begin{array}{c} -\vec{b}_2^* \\ 
                                          \vec{b}_1^* 
                  \end{array}\right), 
\]
it thus follows that $\op{U}$ maps $\vec{c}\equiv \op{J}\vec{a}^*$ onto $\vec{d}\equiv
\op{J}\vec{b}^*$.   Therefore, given two (orthogonal) unit vectors of the form $\vec{a}$
and $\vec{c}$, it is not possible to find a unitary transformation in $Sp(\ell)$ that 
maps these two vectors onto two arbitrary (orthogonal) unit vectors.  Rather, once we 
have chosen the image of $\vec{a}$, the image of $\vec{c}$ is fixed.
\end{example}
This lack of two-point transitivity has serious implications for the action of 
$Sp(\frac{1}{2}N)$, in particular it implies non-transitive action on all kinematical 
equivalence classes of type (\ref{type3}).

\begin{lemma} \label{lemma2}
$Sp(\frac{1}{2}N)$ does not act transitively on kinematical equivalence classes of 
density matrices with at least three distinct eigenvalues, two of which having 
multiplicity one. 
\end{lemma}

\begin{proof}
Any two kinematically equivalent density matrices can be written as 
\[
  \op{\rho}_0=\sum_{n=1}^N w_n \ket{\Psi_n}\bra{\Psi_n}, \quad
  \op{\rho}_1=\sum_{n=1}^N w_n \ket{\Phi_n}\bra{\Phi_n}.
\]
Since there are at least three distinct eigenvalues and two of them have multiplicity
one, we may assume $w_1\neq w_n$ for all $n\neq 1$ and $w_2\neq w_n$ for all $n\neq 2$.
Thus, $\ket{\Psi_n}$ and $\ket{\Phi_n}$ for $n=1,2$ are unique up to phase factors and
any $\op{U}$ such that $\op{\rho}_1=\op{U}\op{\rho}_0\op{U}^\dagger$ must map 
$\ket{\Psi_n}$ onto $\ket{\Phi_n}$ (modulo phase factors) for $n=1,2$, i.e.,
\[
  \op{U}\ket{\Psi_1}=e^{i\phi_1}\ket{\Phi_1}, \quad
  \op{U}\ket{\Psi_2}=e^{i\phi_2}\ket{\Phi_2}, 
\]
However, suppose $\ket{\Psi_1}\doteq\vec{a}$, $\ket{\Psi_2}\doteq\vec{c}$ and 
$\ket{\Phi_1}\doteq \vec{b}$ but $\ket{\Phi_2}\neq e^{i\phi}\vec{d}$, where $\vec{a}$, 
$\vec{b}$, $\vec{c}$ and $\vec{d}$ are as defined in example \ref{example:one}.  This
example then shows that it is impossible to find a $\op{U}\in Sp(\frac{1}{2}N)$ that 
simultaneously maps $\ket{\Psi_1}$ onto $\ket{\Phi_1}$ and $\ket{\Psi_2}$ onto 
$\ket{\Phi_2}$.  Therefore, there does not exist a unitary operator in $Sp(\frac{1}{2}N)$
such that $\op{\rho}_1=\op{U}\op{\rho}_0\op{U}^\dagger$. 
\end{proof}

\begin{lemma} \label{lemma3}
$Sp(\frac{1}{2}N)$ does not act transitively on equivalence classes of density matrices
that have at least one non-zero eigenvalue that occurs with multiplicity greater than
one but less than $N-1$.
\end{lemma}

\begin{proof}
Suppose $w_1$ has multiplicity $N_1$ where $2\le N_1\le N-2$.  If $Sp(\frac{1}{2}N)$ acts
transitively on the selected equivalence class of states then we must be able to map the
$N_1$-dimensional eigenspace $E^{(0)}(w_1)$ for $\op{\rho}_0$ onto the corresponding
eigenspace $E^{(1)}(w_1)$ for $\op{\rho}_1$ by a unitary operator in $Sp(\frac{1}{2}N)$
for any $\op{\rho}_0$ and $\op{\rho}_1$ in the same equivalence class.  However, it is 
easy to see that this is not always possible.  Suppose $E^{(0)}(w_1)$ contains a pair of
vectors of the form $\vec{a}$, $\vec{c}$ as defined above and $E^{(1)}(w_1)$ contains
a vector $\vec{b}$ but the related vector $\vec{d}$ is in the orthogonal complement of
$E^{(1)}(w_1)$.  Then it is impossible to map $E^{(0)}(w_1)$ onto $E^{(1)}(w_1)$ by a
$\op{U}\in Sp(\frac{1}{2}N)$.  Since the orthogonal complement of $E^{(1)}(w_1)$ has at
least dimension two, we can always choose $E^{(1)}(w_1)$ such that $\vec{d}\in E^{(1)}
(w_1)^\perp$.  Hence, $Sp(\frac{1}{2}N)$ does not act transitively on the selected 
equivalence class of states.
\end{proof}
Given any two mixed states $\op{\rho}_0$ and $\op{\rho}_1$ related by $\op{\rho}_1=
\op{U}\op{\rho}_0\op{U}^\dagger$ for some $\op{U}\in Sp(\frac{1}{2}N)\times U(1)$, we
can find a $\tilde{U}\in Sp(\frac{1}{2}N)$ such that $\op{\rho}_1=\tilde{U}\op{\rho}_0
\tilde{U}^\dagger$.  For instance, if $\det\op{U}=e^{i\alpha}$, setting $\tilde{U}=
e^{-i\alpha/N}\op{U}$ produces an operator with $\det(\tilde{U})=1$ that obviously 
satisfies 
\[ 
 \tilde{U}\op{\rho}_0\tilde{U}^\dagger=\op{U}\op{\rho}_0\op{U}^\dagger=\op{\rho}_1.
\]
Thus, $Sp(\frac{1}{2}N)\times U(1)$ acts transitively on a kinematical equivalence 
class $\C$ of density matrices if and only if $Sp(\frac{1}{2}N)$ does.  Combining
this observation with the previous lemmas yields the following theorem.

\begin{theorem}\hfill\null
\begin{itemize}
\item $U(N)$ and $SU(N)$ act transitively on all kinematical equivalence classes.
\item $Sp(\frac{1}{2}N)$ and $Sp(\frac{1}{2}N)\times U(1)$ act transitively on all 
      kinematical equivalence classes of density matrices of type (\ref{type1}) or 
      (\ref{type2}) and \emph{only} those.
\item Any other dynamical Lie group acts transitively only on the trivial kinematical
      equivalence class of completely random ensembles.
\end{itemize}
\end{theorem}

\section{Criteria for reachability of target states}

Having established that the action of the dynamical Lie groups $Sp(\frac{1}{2}N)$ and
$Sp(\frac{1}{2}N)\times U(1)$ is \emph{not} transitive on any kinematical equivalence
class of density matrices of type (\ref{type3}), and that all other dynamical Lie groups
except $U(N)$ and $SU(N)$ act transitively only on the trivial kinematical equivalence 
class of completely random ensembles, the question of identifying states that are 
kinematically but not dynamically equivalent arises.

Since dynamical Lie groups can be very complicated, it would be unrealistic to expect
that simple criteria for dynamical equivalence of states can be derived for arbitrary
dynamical Lie groups.  However, for certain types of dynamical Lie groups of special 
interest, such as $Sp(\frac{1}{2}N)$ [or $Sp(\frac{1}{2}N)\times U(1)$] and $SO(N)$ 
[or $SO(N)\times U(1)$], this is possible, as will be shown in the following.

\subsection{Systems with dynamical Lie group $Sp(\frac{1}{2}N)$ or 
            $Sp(\frac{1}{2}N)\times U(1)$}

To address the problem of finding criteria for dynamical equivalence of states for systems
whose dynamical Lie group $S$ is isomorphic (unitarily equivalent) to $Sp(\frac{1}{2}N)$, 
we recall that any unitary operator $\op{U}\in Sp(\ell)$ satisfies $\op{U}^T\op{J}\op{U}=
\op{J}$ for $\op{J}$ as defined in (\ref{eq:Jsp}).  Thus, any dynamical evolution operator
$\op{U}$ for a system of dimension $N=2\ell$ with dynamical Lie group of type $Sp(\ell)$ 
must satisfy 
\begin{equation}
 \op{U}^T\tilde{J}\op{U}=\tilde{J}
\end{equation}
for a matrix $\tilde{J}$, which is unitarily equivalent to (\ref{eq:Jsp}).%
\footnote{See \ref{appendix:A} for details about how to determine $\tilde{J}$.}
Therefore, we must have 
\[
  \op{U} = \tilde{J}^\dagger \op{U}^* \tilde{J}, \quad
  \op{U}^\dagger = \tilde{J}^\dagger \op{U}^T \tilde{J}.
\]
Two kinematically equivalent states $\op{\rho}_0$ and $\op{\rho}_1$ are thus dynamically
equivalent if and only if there exists a unitary operator $\op{U}$ such that 
\[ 
  \op{\rho}_1=\op{U}\op{\rho}_0\op{U}^\dagger \quad \mbox{and} \quad
  \op{\rho}_1=\tilde{J}^\dagger\op{U}^*\tilde{J}\op{\rho}_0\tilde{J}^\dagger\tilde{U}^T\tilde{J},
\]
or equivalently,
\begin{equation} \label{eq:DE}
  \op{\rho}_1 = \op{U}\op{\rho}_0\op{U}^\dagger \mbox{ and }
  \underbrace{(\tilde{J}\op{\rho}_1\tilde{J}^\dagger)^*}_{\tilde{\rho}_1}
  =\op{U}\underbrace{(\tilde{J}\op{\rho}_0\tilde{J}^\dagger)^*}_{\tilde{\rho}_0}\op{U}^\dagger.
\end{equation}

\begin{example} Let $N=4$ and $S=Sp(2)$ with $\tilde{J}=\op{J}$ as in (\ref{eq:Jsp}).\nobreak
\begin{enumerate}
\item Then $\op{\rho}_0=\mbox{diag}(a,a,b,b)$ ($0\le a,b \le \frac{1}{2}$, $a+b=
      \frac{1}{2}$) and $\op{\rho}_1=\mbox{diag}(a,b,b,a)$ are dynamically equivalent
      since there exists a unitary operator $\op{U}$ such that $\op{\rho}_1=\op{U}
      \op{\rho}_0\op{U}^\dagger$ and any such $\op{U}$ clearly maps $\tilde{\rho}_0=
      \mbox{diag}(b,b,a,a)$ to $\tilde{\rho}_1 =\mbox{diag}(b,a,a,b)$.
\item $\op{\rho}_0$ and $\op{\rho}_2=\mbox{diag}(a,b,a,b)$, on the other hand, are not 
      dynamically equivalent (unless $b=a$) since $\tilde{\rho}_2=\op{\rho}_2$ but 
      $\tilde{\rho}_0\neq\op{\rho}_0$ and there cannot be a unitary operator such that 
      $\op{\rho}_1=\op{U}\op{\rho}_0\op{U}^\dagger=\op{U}\tilde{\rho}_0\op{U}^\dagger$ 
      if $\op{\rho}_0\neq \tilde{\rho}_0$.
\end{enumerate}
This shows that $S=Sp(2)$ divides any kinematical equivalence class of states with two 
distinct eigenvalues of multiplicity $\ell=2$ into at least two disjoint subsets of
dynamically equivalent states.
\end{example}
Sometimes the condition $\op{U}^T\tilde{J}\op{U}=\tilde{J}$ can also be used directly to 
show that two states are not dynamically equivalent.
\begin{example}
Consider again $N=4$ and $S=Sp(2)$ with $\tilde{J}=\op{J}$ as in (\ref{eq:Jsp}) as well as
the initial state $\op{\rho}_0=\mbox{diag}(a,b,c,d)$ where $0\le a,b,c,d\le 1$, $a+b+c+d=1$
and $a,b,c,d$ mutually different.  We can conclude that the state $\op{\rho}_1=\diag(b,a,c,d)$
is not dynamically equivalent to $\op{\rho}_0$ since we would require a unitary operator of 
the form
\[
  \op{U} = \left(\begin{array}{cccc} 0 & e^{i\phi_1} & 0 & 0 \\
                                     e^{i\phi_2} & 0 & 0 & 0 \\
                                     0 & 0 & e^{i\phi_3} & 0 \\
                                     0 & 0 & 0 & e^{i\phi_4} 
                 \end{array} \right)
\]
which does not satisfy $\op{U}^T\op{J}\op{U}=\op{J}$.
\end{example}
Another way of showing that two (kinematically equivalent) density matrices are not 
dynamically equivalent is to prove that (\ref{eq:DE}) cannot have a solution by showing
that the related linear system
\begin{equation} \label{eq:DE2}
  \op{\rho}_1\op{U} - \op{U}\op{\rho}_0 = 0, \quad
  \tilde{\rho}_1 \op{U} - \op{U}\tilde{\rho}_0 = 0
\end{equation}
does not have a solution.  To verify this, we note that the linear system above can be 
rewritten in the form $\A\vec{U}=0$ where $\A$ is a matrix with $2N^2$ rows and $N^2$
columns and $\vec{U}$ is a column vector of length $N^2$.  If the null space of $\A$ is
empty then there is no $\vec{U}$ such that $\A\vec{U}=0$ and hence there is no $N\times
N$ matrix $\op{U}$ that satisfies (\ref{eq:DE2}).  However, note that if the linear
system above \emph{does} have a solution, this does not imply that the states in question
are dynamically equivalent since the solution to the linear equation is in general not 
unitary.

\subsection{Systems with dynamical Lie group $SO(N)$ or $SO(N)\times U(1)$}

From the previous discussion, we know that $SO(N)$ does not act transitively on any
kinematical equivalence class other than the trivial one.  However, we can establish 
criteria for dynamical equivalence of states similar to those for $Sp(\frac{1}{2}N)$ 
by noting that any unitary operator $\op{U}\in SO(N)$ must satisfy $\op{U}^T\op{J}\op{U}
=\op{J}$ for $\op{J}$ as in (\ref{eq:Jso}).  Therefore, two kinematically equivalent 
states $\op{\rho}_0$ and $\op{\rho}_1$ are dynamically equivalent under the action of 
a dynamical Lie group $S$ which is unitarily equivalent to $SO(N)$, if there exists a 
unitary operator $\op{U}$ such that
\begin{equation}
  \op{\rho}_1 = \op{U}\op{\rho}_0\op{U}^\dagger \mbox{ and }
  \underbrace{(\tilde{J}\op{\rho}_1\tilde{J}^\dagger)^*}_{\tilde{\rho}_1}
  =\op{U}\underbrace{(\tilde{J}\op{\rho}_0\tilde{J}^\dagger)^*}_{\tilde{\rho}_0}\op{U}^\dagger.
\end{equation}
with $\tilde{J}$ unitarily equivalent to (\ref{eq:Jso}), and determined as described in 
\ref{appendix:A}.
\begin{example}
Consider a system with $N=5$ and Hamiltonian $\op{H}=\op{H}_0+f(t)\op{H}_1$ where
\[
  \op{H}_0=\left(\begin{array}{ccccc} -2 & 0 & 0 & 0 & 0 \\      
                                       0 &-1 & 0 & 0 & 0 \\
                                       0 & 0 & 0 & 0 & 0 \\
                                       0 & 0 & 0 & 1 & 0 \\
                                       0 & 0 & 0 & 0 & 2
                 \end{array} \right), \qquad 
  \op{H}_1=\left(\begin{array}{ccccc}  0 & 1 & 0 & 0 & 0 \\      
                                       1 & 0 & 1 & 0 & 0 \\
                                       0 & 1 & 0 & 1 & 0 \\
                                       0 & 0 & 1 & 0 & 1 \\
                                       0 & 0 & 0 & 1 & 0
                 \end{array} \right).
\]
It can be verified using the algorithm described in \cite{PRA63n063410} that the Lie
algebra of this system has dimension $10$, which is equal to the dimension of $so(5)$.
Using the technique described in \ref{appendix:A}, we find that both of the generators 
$\rmi\op{H}_0$ and $\rmi\op{H}_1$ of the Lie algebra satisfy $\op{x}^T\tilde{J}+\tilde{J}
\op{x}=0$ for 
\[
  \tilde{J} = \left(\begin{array}{ccccc}  0 & 0 & 0 & 0 & 1 \\      
                                       0 & 0 & 0 &-1 & 0 \\
                                       0 & 0 & 1 & 0 & 0 \\
                                       0 &-1 & 0 & 0 & 0 \\
                                       1 & 0 & 0 & 0 & 0
                 \end{array} \right),  
\]
which is unitarily equivalent to the standard $\op{J}$ for $so(5)$.  We can thus 
conclude that its dynamical Lie algebra is $so(5)$ and its dynamical Lie group is
$SO(5)$.  Furthermore, note that the two pure states
\[
  \op{\rho}_0=\left(\begin{array}{ccccc}   1 & 0 & 0 & 0 & 0 \\      
                                           0 & 0 & 0 & 0 & 0 \\
                                           0 & 0 & 0 & 0 & 0 \\
                                           0 & 0 & 0 & 0 & 0 \\
                                           0 & 0 & 0 & 0 & 0
                 \end{array} \right), \qquad
  \op{\rho}_1=\left(\begin{array}{ccccc} 0.5 & 0 & 0 & 0 & 0.5 \\      
                                           0 & 0 & 0 & 0 & 0 \\
                                           0 & 0 & 0 & 0 & 0 \\
                                           0 & 0 & 0 & 0 & 0 \\
                                         0.5 & 0 & 0 & 0 & 0.5
                 \end{array} \right)
\]
are not dynamically equivalent since $(\tilde{J}\op{\rho}_1\tilde{J}^\dagger)^*=\op{\rho}_1$
but $(\tilde{J}\op{\rho}_0\tilde{J}^\dagger)^*\neq\op{\rho}_0$ and it is thus impossible to 
find a unitary transformation such that $\op{U}\op{\rho}_0\op{U}^\dagger=\op{\rho}_1
=\op{U}(\tilde{J}\op{\rho}_0\tilde{J}^\dagger)^*\op{U}^\dagger$.
\end{example}
%
\section{Conclusion}

The question of dynamical equivalence of kinematically equivalent quantum states has 
been been addressed by studying the action of the dynamical Lie group of the system
on the kinematical equivalence classes.  For systems whose dynamical Lie group is 
unitarily equivalent to either $Sp(\frac{1}{2}N)$ or $SO(N)$, explicit criteria for
dynamical reachability / equivalence of states have been given, and their application
illustrated with several examples.

Furthermore, we have provided a classification of density matrices according to their 
eigenvalues, which divides mixed quantum states into three main types: (i) completely 
random ensembles, (ii) pure-state-like ensembles, and (iii) general ensembles.  We have
also proved that the dynamical Lie group $Sp(\frac{1}{2}N)$ acts transitively on all 
equivalence classes of quantum states of type (\ref{type1}) and (\ref{type2}), but 
\emph{only} those.  

Although it is known that a pure-state controllable system whose dynamical Lie group 
$S$ is isomorphic to $Sp(\frac{1}{2}N)$ is \emph{not} density matrix controllable in 
general \cite{qph0106128}, this result shows that there are more than just a few 
examples of kinematically equivalent density matrices that are not dynamically reachable
from one another in this case.  In fact, the action of $S$ is \emph{not} transitive on 
\emph{almost all} kinematical equivalence classes.  This is in marked contrast to the
action of $S$ for a density matrix controllable system, which is transitive on \emph{all}
kinematical equivalence classes, as well as the action of $S$ for a non-controllable 
system, which is transitive only on the trivial kinematical equivalence class of 
completely random ensembles.  

\ack  

SGS acknowledges the hospitality and financial support of the Department of Mathematics
and the Institute of Theoretical Science at the University of Oregon, where most of 
this work was completed.  AIS acknowledges the hospitality of the Laboratoire de 
Physique Th\'eorique des Liquides, University of Paris IV, where he is currently a 
visiting faculty member.

\appendix
\section{Finding $\op{J}$ for dynamical Lie groups of type 
          $Sp(\frac{1}{2}N)$ or $SO(N)$} \label{appendix:A}

For the results of the previous sections to be truly useful, we must also address the
question of how to determine the $\tilde{J}$ matrix of a given system.  To this end, note
that the elements of the dynamical Lie algebra $L$ associated with the dynamical groups 
$Sp(\frac{1}{2}N)$ and $SO(N)$ must satisfy a relation similar to the one satisfied by
the elements of the group, namely any $\op{x}\in L$ must satisfy
\begin{equation} \label{eq:Jcond}
  \op{x}^T \tilde{J}+ \tilde{J}\op{x} = 0,
\end{equation}
where $\tilde{J}$ is the same as for the related group.  Thus, given a system with total 
Hamiltonian (\ref{eq:H}), this implies in particular that the generators $\rmi\op{H}_m$
of the dynamical Lie algebra must satisfy (\ref{eq:Jcond}). 

Equation (\ref{eq:Jcond}) can be written as a system of linear equations of the form
\[
  \L_m \vec{J} = 0, \quad 0\le m\le M,
\]
where $\L_m$ is a square matrix of dimension $N^2$ determined by the generators 
$\rmi\op{H}_m$ and $\vec{J}$ is a column vector of length $N^2$.  The solutions 
$\vec{J}$ of the above matrix equation can be found by computing the null space 
of the operator 
\[
  \left( \begin{array}{c} \tilde{\L}_0 \\ \vdots \\ \tilde{\L}_M \end{array} \right).
\]
If the dynamical Lie group is of type $Sp(\frac{1}{2}N)$ or $SO(N)$ then the nullspace
contains a single element $\vec{J}$, which can be rearranged into a square matrix whose
eigenvalues agree with whose of the standard $\op{J}$ for the group defined above.
That is, concretely, 
\begin{itemize}
\item if $N=2\ell$ and $\tilde{J}$ has two distinct eigenvalues $+\rmi$ and $-\rmi$, 
      both of which occur with multiplicity $\ell$ then the dynamical Lie group is 
      $Sp(\ell)$; 
\item if $N=2\ell$ and $\tilde{J}$ has two distinct eigenvalues $+1$ and $-1$, both of 
      which occur with multiplicity $\ell$ then the dynamical Lie group is 
      $SO(2\ell)$; 
\item if $N=2\ell+1$ and $\tilde{J}$ has two distinct eigenvalues $+1$ and $-1$, 
      occurring with multiplicity $\ell+1$ and $\ell$, respectively, then the dynamical
      Lie group is $SO(2\ell+1)$; 
\end{itemize}
Hence, the algorithm not only determines $\tilde{J}$ but it also allows us to decide 
whether the dynamical Lie group is of type $Sp(\frac{1}{2}N)$ or $SO(N)$.  

Note that the dynamical Lie group $S$ can only be $Sp(\frac{1}{2}N)$ or $SO(N)$ if all 
the partial Hamiltonians $\op{H}_m$ of the system have zero trace.  However, if any of
the partial Hamiltonians $\op{H}_m$ has non-zero trace then the dynamical Lie group of
the system can still be $Sp(\frac{1}{2}N)\times U(1)$ or $SO(N)\times U(1)$.  To deal 
with this situation, we note that $S \simeq Sp(\frac{1}{2}N)\times U(1)$ or $S\simeq
SO(N)\times U(1)$ is possible only if the generators
\begin{equation}
  \op{x}_m=\rmi\op{H}_m -\frac{\rmi}{N}\Tr(\op{H}_m)\op{I}_N, \quad 0\le m\le M
\end{equation}
of the related trace-zero Lie algebra $L'$ satisfy (\ref{eq:Jcond}) for $0\le m\le M$
and we can thus proceed as above to determine $\tilde{J}$.

\section{Comparison of Theorem 1 with Theorem 6 in \cite{qph0106128}}
\label{appendix:B} 

To demonstrate the difficulty in using theorem 6 in \cite{qph0106128} to verify whether
the dynamical Lie group $S$ of a system acts transitively on an equivalence class of
density operators, we shall consider a simple example.

Assume the dynamical Lie group of the system is $Sp(2) \subset U(4)$.  According to 
theorem 1 above, $Sp(2)$ does \emph{not} act transitively on the kinematical equivalence
class represented by $\op{\rho}=\mbox{diag}(a,a,b,b)$ with $0\le a,b\le \frac{1}{2}$ and
$a+b=\frac{1}{2}$ since $\op{\rho}$ is of type (\ref{type3}).

To show that the action is not transitive using theorem 6 in \cite{qph0106128}, we note 
first that $\dim U(4)=16$ and $\dim Sp(2)=10$.  Thus, the left hand side in 
(\ref{eq:dimFormula}) is $\dim U(4)-\dim Sp(2)=6$.  

To compute the right hand side, we need to determine the centralizer $\C_\op{\rho}$ of
$\op{\rho}$.  Noting that 
\[
  \op{\rho}= \left( \begin{array}{cc} a \op{I}_2 & 0 \\ 0 & b \op{I}_2 \end{array}\right),
\]
where $\op{I}_2$ is the indentity matrix in dimension 2, we see that $\op{\rho}$ commutes
with every unitary matrix of the form
\[
  \op{U}_C= \left( \begin{array}{cc} \op{A}_1 & 0 \\ 0 & \op{A}_2 \end{array}\right),
\]
where $\op{A}_1$ and $\op{A}_2$ are arbitrary unitary matrices in $U(2)$.  Thus, the
centralizer of $\op{\rho}$ is $U(2)\times U(2)$ and its dimension is $4+4=8$.

To compute the intersection of $\C_\op{\rho}$ with $S=Sp(2)$, we recall that any matrix 
in $Sp(2)$ must preserve $\op{J}$ as defined in (\ref{eq:Jsp}).  Concretely, this means
$\op{U}_C^T\op{J}\op{U}_C =\op{J}$, i.e.,
\[
 \op{U}_C^T \op{J} \op{U}_C = 
\left( \begin{array}{cc} 
        0                   & \op{A}_1^T\op{A}_2 \\ 
        -\op{A}_2^T\op{A}_1 & 0 
       \end{array}\right)
= \left(\begin{array}{cc} 
         0                  & \op{I}_2 \\ 
         -\op{I}_2          & 0 
        \end{array}\right).
\]
Thus, we must have $\op{A}_1^T\op{A}_2 = \op{I}_2$.  Noting that $\op{A}_1$ and $\op{A}_2$
are unitary, this is only possible if $\op{A}_1=\op{A}_2^*$, i.e., if $\op{A}_1$ is the 
complex conjugate of $\op{A}_2$, since $(\op{A}_2^*)^T\op{A}_2=\op{A}_2^\dagger\op{A}_2
=\op{I}$.  Hence, the intersection of the centralizer $\C_\op{\rho}$ with $S=Sp(2)$ is
$U(2)$, which has dimension $4$.  Hence, $\dim\C_\op{\rho}-\dim (\C_\op{\rho}\cap Sp(2))
=8-4=4\neq 6$, i.e., the left and right hand side in (\ref{eq:dimFormula}) are \emph{not}
equal.  Thus we have shown using theorem 6 that the action of $Sp(2)$ on the kinematical
equivalence class of $\op{\rho}$ is not transitive.

\section*{References}
\bibliography{books,papers2000,papers9599,papers9094,papers8089,papers--79}
\end{document}